\documentclass[article]{IEEEtran}

\usepackage{cite,graphicx,amsmath,psfrag,bm}
\usepackage[usenames,dvipsnames]{xcolor}
\usepackage{mathabx}
\usepackage{subfig}
\usepackage{cancel}
\usepackage{epstopdf}
\usepackage{pstool}
\usepackage{slashbox}
\usepackage{multirow}
\usepackage{booktabs}

\newcommand{\ve}[1]{\boldsymbol{#1}}
\newcommand{\te}[1]{\overline{\overline{#1}}}

\makeatletter\def\CT{\def\@captype{figure}}\makeatother

\begin{document}

\title{Synthesis and Applications \\ of Birefringent Metasurfaces}

\author{Karim~Achouri~\IEEEmembership{Student,~IEEE},
        Guillaume~Lavigne,
        and~Christophe~Caloz,~\IEEEmembership{Fellow,~IEEE}

\thanks{K. Achouri, G. Lavigne and C. Caloz are with the Department
of Electrical Engineering, $\acute{\mathrm{E}}$cole polytechnique de Montr$\acute{\mathrm{e}}$al, Montr$\acute{\mathrm{e}}$al,
QC, H3T 1J4 Canada (e-mail: karim.achouri@polymtl.ca).}}

\maketitle

\begin{abstract}
Birefringent metasurfaces are two-dimensional structures capable of independently controlling the amplitude, phase and polarization of orthogonally polarized incident waves. In this work, we propose a in-depth discussion on the mathematical synthesis of such metasurfaces. We compare two methods, one that is rigorous and based on the exact electromagnetic fields involved in the transformation and one that is based on approximate reflection and transmission coefficients. We next validate the synthesis technique in metasurfaces performing the operations of half- and quarter-wave plates, polarization beam splitting and orbital angular momentum multiplexing.
\end{abstract}

\IEEEpeerreviewmaketitle


\section{Introduction}

Birefringence, also called double refraction, is the physical property of an anisotropic medium to exhibit an angle dependent refractive index~\cite{saleh2007fundamentals}. This phenomenon, first observed in crystals more than 300 years ago~\cite{Bartho}, has already lead to the realization of several major optical components and applications. More recently, metasurfaces~\cite{yu2014flat,lin2014dielectric,Pfeiffer2013a,023902,kim2014optical,achouri2014general}, the two dimensional counterparts of metamaterials, have seen an important rise of interest due to their rich potential in the transformation of electromagnetic fields. Combined with the orthogonality property of $x$ and $y$ polarized waves, birefringent structures, and especially metasurfaces, have the capability to independently control the amplitude, phase and polarization of two orthogonal electromagnetic waves, leading to a wealth of possible applications at optical and microwave frequencies.

In this work, we propose an in-depth discussion on the synthesis of birefringent metasurfaces. This discussion is based on the general bianisotropic metasurface synthesis technique developed in~\cite{achouri2014general}. We compare two different synthesis methods, one that is rigorous and one that is based on paraxial approximate transmission and reflection coefficients. We also report the implementation details of several metasurfaces that were only briefly presented in~\cite{art1,AchouriEPJAM}. These metasurfaces perform the operations of half-wave plate~\cite{ding2015broadband,pors2013broadband,lin2014dielectric}, quarter-wave plate~\cite{yu2012broadband,zhao2011manipulating}, polarization beam splitting~\cite{pors2013gap,farmahini2013birefringent,lee2014semiconductor} and orbital angular momentum~\cite{PhysRevApplied.2.044012,karimi2014generating,capasso1,chen2015creating,wang2015ultra}
 multiplexing.

The paper is organized as follows. In the second section, the mathematical synthesis as well as the physical realization of metasurfaces are addressed. Two different approaches for the mathematical synthesis are discussed and compared. In the third section, we present the implementation of the four metasurfaces introduced above.

\section{Metasurface Design}

\subsection{Mathematical Synthesis}

The metasurface synthesis technique in~\cite{achouri2014general} stems from the continuity relations, initially derived by Idemen~\cite{Idemen1973}, which express the discontinuities of the electromagnetic fields in the presence of a spatial discontinuity, such as a metasurface. In the simple case of a monoanisotropic (zero magnetoelectric coupling coefficients) birefringent metasurface given in terms of its transverse diagonal susceptibility tensors, $\te{\chi}_\text{ee}$ and $\te{\chi}_\text{mm}$, the continuity relations read
\begin{subequations}
\label{eq:BC_plane}
\begin{align}
\hat{z}\times\Delta\ve{H}
&=j\omega\epsilon_0\te{\chi}_\text{ee}\cdot\ve{E}_\text{av},\label{eq:BC_plane_1}\\
\Delta\ve{E}\times\hat{z}
&=j\omega\mu_0 \te{\chi}_\text{mm}\cdot\ve{H}_\text{av},\label{eq:BC_plane_2}
\end{align}
\end{subequations}
where it is assumed that the metasurface lies in the $x-y$ plane at $z=0$. In~\eqref{eq:BC_plane}, the differences of the electric and magnetic fields between both sides of the metasurface (expressed by the operator $\Delta$) are related to the metasurface susceptibilities excited by the average electric and magnetic fields (denoted by the subscript `av'). The system of~\eqref{eq:BC_plane} can be easily solved to express the susceptibilities in terms of the specified fields. Due to the orthogonality between $x$ and $y$ polarizations, the solutions split into two independent sets which are respectively given by
\begin{subequations}
\label{eq:X_diag}
\begin{equation}
\chi_{\text{ee}}^{xx}=\frac{-\Delta H_{y}}{j\omega\epsilon_0  E_{x,\text{av}}},\qquad
\chi_{\text{mm}}^{yy}=\frac{-\Delta E_{x}}{j\omega\mu_0  H_{y,\text{av}}},\label{eq:Xx}
\end{equation}
\begin{equation}
\chi_{\text{ee}}^{yy}=\frac{\Delta H_{x}}{j\omega\epsilon_0  E_{y,\text{av}}},\qquad
\chi_{\text{mm}}^{xx}=\frac{\Delta E_{y}}{j\omega\mu_0  H_{x,\text{av}}}.\label{eq:Xy}
\end{equation}
\end{subequations}
At this stage, the metasurface is completely defined by the susceptibilities in~\eqref{eq:X_diag} and performs the required transformation between the incident, reflected and transmitted waves~\cite{achouri2014general,AchouriEPJAM}. The birefringent operation leverages the property of orthogonality between~\eqref{eq:Xx} and~\eqref{eq:Xy}.

In order to realize the metasurface, one has to find the appropriate shapes of the scattering particles. Therefore, it is convenient to find the transmission and reflection coefficients of the metasurface which could then be related to the transmission and reflection coefficients of each scattering particle obtained via full-wave simulations. To obtain these coefficients, it is assumed that the metasurface is illuminated by a normally incident plane wave and that it reflects and transmits normally propagating plane waves (either $x$ or $y$ polarized). The corresponding electric and magnetic fields are thus, respectively, given by $E_\text{i}=e^{-jkz}, E_\text{t}=Te^{-jkz}$ and $E_\text{r}=Re^{jkz}$, and $H_\text{i}=e^{-jkz}/\eta_0, H_\text{t}=Te^{-jkz}/\eta_0$ and $H_\text{r}=-Re^{jkz}/\eta_0$. These fields are then substituted into relations~\eqref{eq:X_diag}, which are then solved for the coefficients $T$ and $R$, which read~\cite{achouri2014general}
\begin{subequations}
\label{eq:RTcoef}
\begin{align}
T&=\frac{4+\chi_{\text{ee}}\chi_{\text{mm}}k_0^2}{(2+jk_0\chi_{\text{ee}})(2+jk_0\chi_{\text{mm}})},\\
R&=\frac{2jk_0\left(\chi_{\text{mm}}-\chi_{\text{ee}}\right)}{\left(2+jk_0\chi_{\text{ee}}\right)\left(2+jk_0\chi_{\text{mm}}\right)},
\label{eq:RTcoef2}\end{align}
\end{subequations}
where $k_0$ is the free space wave number. Using the susceptibilities from~\eqref{eq:Xx} or~\eqref{eq:Xy} will yield the transmission and reflection coefficients for $x$ or $y$ polarizations, respectively. Since a monoanisotropic metasurface is necessarily symmetric with respect to the $z$ direction~\cite{asadchy2016metasurfaces}, the reflection coefficients of our metasurface are the same from both sides of the structure.

Relations~\eqref{eq:RTcoef} can be reversed to express the susceptibilities in terms of the transmission and reflection coefficients as
\begin{subequations}
\label{eq:chi_Sparam}
\begin{align}
\chi_{\text{ee}}&=\frac{2j\left(T+R-1\right)}{k_0\left(T+R+1\right)},\\
\chi_{\text{mm}}&=\frac{2j\left(T-R-1\right)}{k_0\left(T-R+1\right)},
\end{align}
\end{subequations}
where $\chi_{\text{ee}}^{xx}$ and $\chi_{\text{mm}}^{yy}$ are found assuming that $T$ and $R$ are the coefficients of $x$-polarized waves. Alternatively, $\chi_{\text{ee}}^{yy}$ and $\chi_{\text{mm}}^{xx}$ are found when $T$ and $R$ are the coefficients of $y$-polarized waves. Although Eqs.~\eqref{eq:X_diag} can be rigorously used to synthesize metasurfaces, relations~\eqref{eq:chi_Sparam} suggest an alternative synthesis technique which would consists in specifying the required transformation in terms of transmission and reflection coefficients instead of the tangential electromagnetic fields, as would be done in~\eqref{eq:X_diag}.\\

In what follows, we will compute the responses of metasurfaces synthesized using the methods based on relations~\eqref{eq:X_diag} and~\eqref{eq:chi_Sparam} and compare them.  Let us consider a reflection-less metasurface that refracts at $45^\circ$ a normally incident $x$-polarized plane wave. The electromagnetic fields of the incident wave are, at $z=0$, given by $E_{i}^x=1, H_{i}^y=1/\eta_0$, while the fields of the transmitted wave are $E_{t}^x=\sqrt{2}/2e^{-jk_0\sqrt{2}/2x}, H_{t}^y=e^{-jk_0\sqrt{2}/2x}/\eta_0$. The first metasurface synthesis method consists in substituting these fields into~\eqref{eq:Xx}, which results in the following susceptibilities
\begin{subequations}
\label{eq:Sin1}
\begin{align}
\chi_{\text{ee}}^{xx}&=\frac{4j}{k_0}\left (\frac{e^{-jk_0\sqrt{2}/2x}-1}{2+\sqrt{2}e^{-jk_0\sqrt{2}/2x}}\right ),\\
\chi_{\text{mm}}^{xx}&=\frac{j}{k_0}\left (\frac{\sqrt{2}e^{-jk_0\sqrt{2}/2x}-2}{1+e^{-jk_0\sqrt{2}/2x}}\right ).
\end{align}
\end{subequations}
The second synthesis method, based on relations~\eqref{eq:chi_Sparam}, seems a priori unsuitable for such a kind of transformation (i.e. refraction) since relations~\eqref{eq:RTcoef} and~\eqref{eq:chi_Sparam} were obtained assuming that all waves impinging on or scattered by the metasurface are propagating normally to the structure, which is obviously contradictory with the concept of refraction. Indeed, this second synthesis technique rigorously applies \emph{only} to normally propagating waves, but it may also be used as an approximation to synthesize refractive metasurfaces in the case of small refraction angles, i.e. paraxial approximation, as will be shown next. In fact, this metasurface synthesis technique allows one to obtain the material properties, through relations~\eqref{eq:chi_Sparam}, from the transmission and reflection coefficients that would be initially defined using the complex amplitude transmittance method~\cite{saleh2007fundamentals} (in the case of zero reflection) or, more generally, the momentum transformation technique introduced in~\cite{Salem2013c}.

\begin{figure}[h!]
\centering
\includegraphics[width=1\columnwidth]{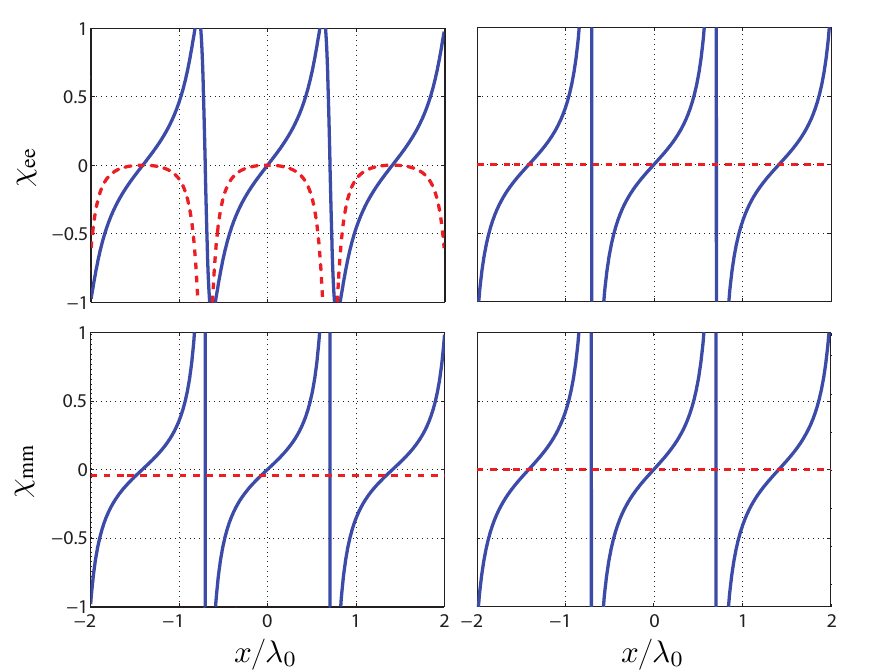}
\caption{Susceptibilities of a metasurface refracting a normally incident plane wave under $45^\circ$. The solid blue lines correspond to real parts while the dashed red lines correspond to imaginary parts. The susceptibilities in the left are obtained with relations~\eqref{eq:X_diag}, while those in the right are obtained from~\eqref{eq:chi_Sparam}.}
\label{Fig:X}
\end{figure}

In the case of a metasurface illuminated by a normally impinging incident wave, the condition of zero reflection, that may be deduced from~\eqref{eq:RTcoef2} with $R=0$, is that $\chi_{\text{ee}}=\chi_{\text{mm}}=\chi$. This reduces equations~\eqref{eq:RTcoef} and~\eqref{eq:chi_Sparam} to
\begin{equation}
T=\frac{2-jk_0\chi}{2+jk_0\chi}
\end{equation}
and
\begin{equation}
\label{eq:X_T}
\chi=\frac{2j\left(T-1\right)}{k_0\left(T+1\right)},
\end{equation}
respectively. Because this metasurface is assumed to be reflection-less, the complex amplitude transmittance method can be used to define the parameter $T$ in~\eqref{eq:X_T}, that is to say
\begin{equation}
\label{eq:T_opt}
 T =\frac{\Psi_t}{\Psi_i},
\end{equation}
where $\Psi_i$ and $\Psi_t$ are the phase profiles, projected on the metasurface plane, of the incident and the transmitted waves, respectively. Since the incident wave is normally impinging on the metasurface, we have that $\Psi_i = 1$. The transmission coefficient is then simply defined by $T =\Psi_t=e^{-jk_0\sqrt{2}/2x}$, which transforms~\eqref{eq:X_T} to
\begin{equation}
\label{eq:X_approx}
\chi=\frac{2}{k_0}\tan{\left(\frac{k_0 x}{2\sqrt{2}}\right)}.
\end{equation}
Now, let us compare the susceptibilities in~\eqref{eq:Sin1}, obtained from the first synthesis method, which is rigorous, and the susceptibilities in~\eqref{eq:X_approx}, derived from the second method, which is approximate. The real and imaginary parts of the electric and magnetic susceptibilities are plotted at the top and bottom of Fig.~\ref{Fig:X}, respectively. The left-hand side of the figure corresponds to the first method while the right-hand side corresponds to the second method. They are two main differences between the two methods. Firstly, the electric and magnetic susceptibilities are different in~\eqref{eq:Sin1} whereas they are equal in~\eqref{eq:X_approx}. Secondly, the susceptibilities in~\eqref{eq:Sin1} are complex with negative imaginary parts corresponding to absorption whereas the susceptibilities in~\eqref{eq:X_approx} are purely real and thus correspond to a lossless and passive structure.
The reason for the imaginary parts in~\eqref{eq:Sin1}, and the resulting absorption, is the unequal normal power flow between the incident wave and the transmitted wave~\cite{7274678,asadchy2016metasurfaces}. Indeed, the transmitted wave, which propagates under $45^\circ$, has a lower normal transmitted power than the normally incident wave. This translates into a reduced transmission efficiency where the excess energy of the incident wave is absorbed by the metasurface.

\begin{figure}[h!]
\centering
\includegraphics[width=1\columnwidth]{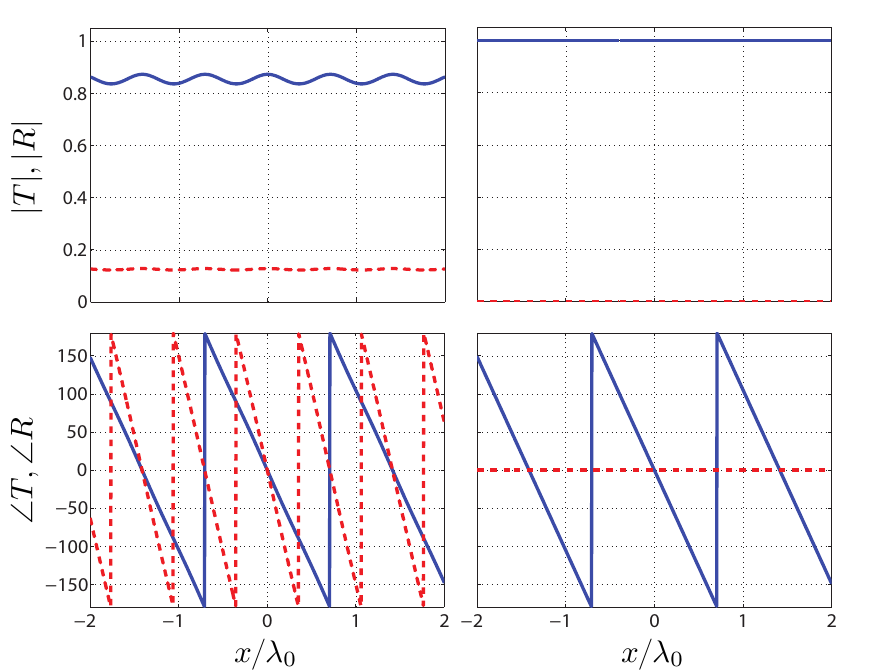}
\caption{Transmission (blue solid line) and reflection (red dashed line) coefficients for the metasurfaces given by the susceptibilities in Fig.~\ref{Fig:X}. The top and bottom plots correspond to the amplitude and phase of these coefficients, respectively. The plots on the left are obtained with relations~\eqref{eq:X_diag}, while the ones on the right are obtained from~\eqref{eq:chi_Sparam}.}
\label{Fig:TR}
\end{figure}

Another interesting comparison to establish between the two methods is the differences of the transmission and reflection coefficients in~\eqref{eq:RTcoef} and~\eqref{eq:T_opt}. Substituting the susceptibilities found in~\eqref{eq:Sin1} into~\eqref{eq:RTcoef} yields the transmission and reflection coefficients, for the first synthesis method, that are plotted on the left-hand side of Fig.~\ref{Fig:TR} in solid blue lines and in dashed red lines, respectively. The transmission and reflection coefficients of the second synthesis method are simply $T=e^{-jk_0\sqrt{2}/2x}$ and $R=0$, and are plotted on the right-hand side of Fig.~\ref{Fig:TR}.
It appears that the metasurface synthesized with the first method may be seen as an equivalent amplitude and phase grating in transmission and reflection, while the other metasurface is a simple transmission phase gradient structure. The non-zero reflection coefficient that is plotted in Fig.~\ref{Fig:TR} (top-left) seems, a priori, contradictory with the prescription of zero reflection specified to obtain relations~\eqref{eq:Sin1}. In fact, no propagating reflected wave is produced by the metasurface because the $k$-vector of the reflection phase (bottom-left in Fig.~\ref{Fig:TR}), defined as $k_r = 2\pi/P_r$, where $P_r$ is the period of the reflection phase, is larger than the free space wave number, $k_0$. This means that the reflected wave is an evanescent wave and thus does not propagate. Moreover, the non ideal transmission efficiency that was discussed above and that is responsible for the imaginary parts of the susceptibilities in~\eqref{eq:Sin1} also contributes to the non-zero reflection coefficient in Fig.~\ref{Fig:TR} and the loss evidenced by the fact that $|T|^2+|R|^2<1$.
\begin{figure}[h!]
\centering
\includegraphics[width=1\columnwidth]{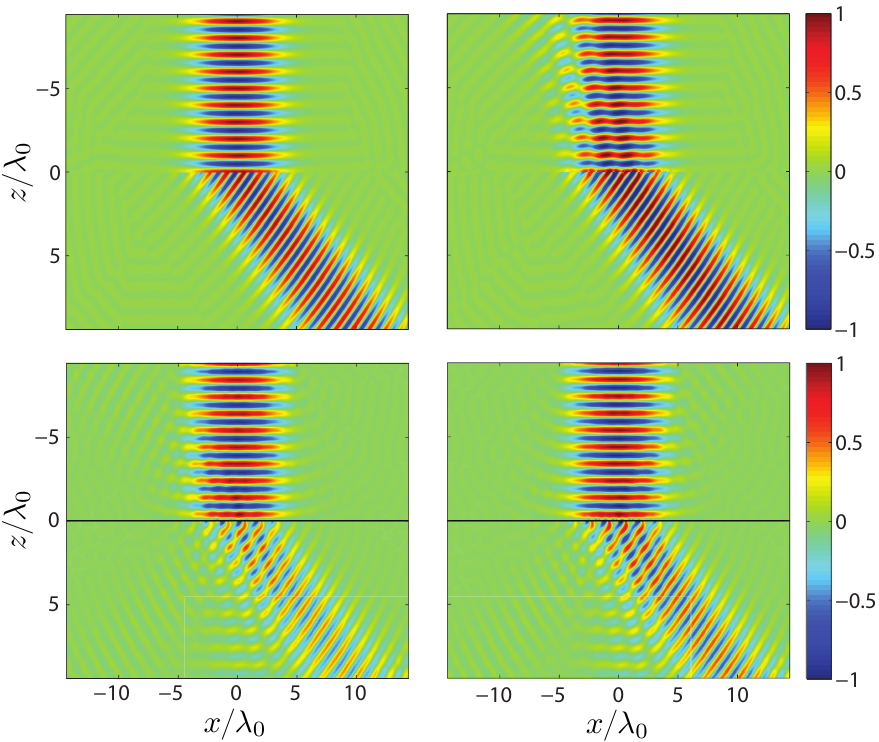}
\caption{Full-wave simulated real part of the $H_z\eta_0$ component. The left side corresponds to the first synthesis method with susceptibilities as in~\eqref{eq:Sin1} while the right side corresponds to the second synthesis method with susceptibilities as in~\eqref{eq:X_approx}. The results in the top row were obtained using an FDFD code and, in the bottom row, using COMSOL.}
\label{Fig:Sims}
\end{figure}

Finally, let us see how these two synthesis techniques compare by performing full-wave simulations. We have made two slightly different kinds of simulations. One using an home made Finite-Difference Frequency Domain (FDFD) code~\cite{vahabzadeh2016simulation} that simulates an exactly zero-thickness metasurface. The other simulations were made using COMSOL by assuming a thin metasurface of thickness $d=\lambda_0/100$. Note that, in COMSOL, the susceptibilities were introduced by assigning relative permittivity and permeability to a slab which are respectively defined by $\te{\epsilon}_r=\te{I}+\te{\chi}_\text{ee}/d$ and $\te{\mu}_r=\te{I}+\te{\chi}_\text{mm}/d$ where the division by $d$ dilutes the effect of the susceptibilities over the thickness of the slab~\cite{Idemen1973}. Simulation results showing the real part of $H_z$ are plotted in Fig.~\ref{Fig:Sims}, where the top and bottom rows correspond to the FDFD simulations and COMSOL simulations, respectively. The left- and the right-hand sides correspond to the first and second synthesis techniques, respectively.

As can be seen, the best result (Fig.~\ref{Fig:Sims} top left) was obtained with the FDFD code and using the rigorous relations in~\eqref{eq:X_diag} (first synthesis technique). This is not surprising because, in this simulation, the metasurface has exactly zero-thickness and, therefore, it can be rigorously described by the continuity equations~\eqref{eq:BC_plane}. The diffraction efficiency, defined as the transmitted power density at $45^\circ$ divided by the incident power density, is about $99\%$ for the first method and $97\%$ for the second one which, surprisingly, works fairly well except for undesired reflection. The simulations with COMSOL give worse results due to the thickness of the metasurface. For both synthesis techniques, several diffraction orders appear, either in reflection or in transmission. Moreover, due to the thickness, some modes are trapped (guided modes) inside the metasurface, which contributes to further lower the diffraction efficiency into the desired direction to $26\%$ and $40\%$ for the first and second methods, respectively. The rigorous method gives, in that case, worse results than the approximate method which might be explained by the fact that the susceptibilities in~\eqref{eq:X_diag} are lossy and thus the wave is more attenuated by propagating through the thickness of the metasurface whereas the metasurface obtained with the second method is not lossy at all.

To conclude this section, it must be noted that, while the first synthesis technique is the most rigorous one and gives the best results in FDFD simulation, it remains much more complicated to implement than the second method. This is because the physical realization of these metasurfaces necessarily requires a mapping between the susceptibilities, given in Fig.~\ref{Fig:X}, and the scattering parameters, given in Fig.~\ref{Fig:TR}. And, as can be seen from the scattering parameters, the realization of the metasurface synthesized with the first method would require implementing non-uniform reflection and transmission coefficients that present different phase gradients, moreover this metasurface would also be lossy. Compare this now to the metasurface synthesized with the second method and that presents a uniform unity transmission coefficient and a phase gradient, it is clear that this second method is much easier to realize and considering the excellent results in FDFD simulation, it is the one that is usually preferred for the realization of most metasurfaces. The metasurfaces presented in the following section of this paper are synthesized based on the second synthesis technique.

\subsection{Physical Realization}

In order to fabricate the metasurfaces, the required scattering parameters are discretized with subwavelength resolution. At each lattice site, a scattering particle (or unit cell) is realized such that it exhibits the required scattering behavior. The unit cells are simulated in a commercial software and assuming periodic boundary conditions. The shape of the unit cells are optimized such that the scattering matrices obtained by simulation correspond to the transmission and reflection coefficients given in~\eqref{eq:RTcoef}.

To implement each unit cell, we have used a cascade of three metallic layers (with identical outer layers) held together by two dielectric spacers. This type of unit cells has been shown to present full transmission (assuming lossless material) and a complete $360^\circ$-phase coverage~\cite{PhysRevApplied.2.044011}.

\begin{figure}[h!]
\begin{center}
\includegraphics[width=0.8\columnwidth]{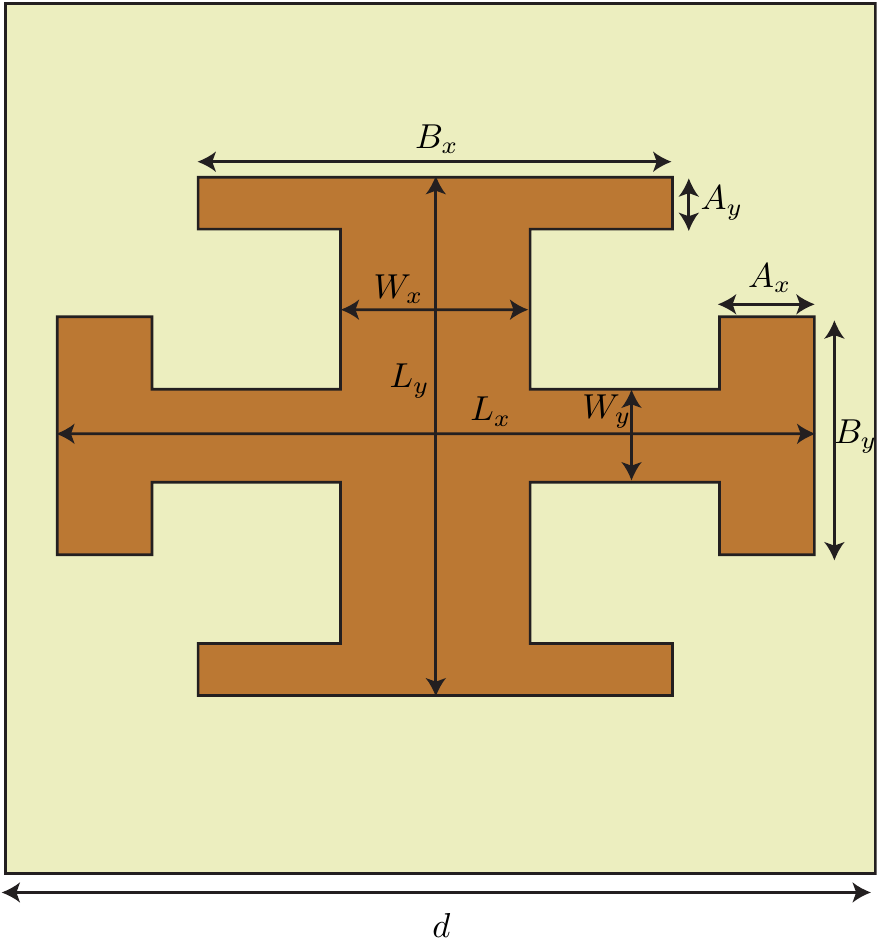}
\caption{Generic metallic layer used to realize the metasurface unit cells.}
\label{Fig:metallicUnitCell}
\end{center}
\end{figure}

Each metallic layer of the unit cell consists in a Jerusalem cross, as shown in Fig.~\ref{Fig:metallicUnitCell} with all its variable dimensions. All the metasurfaces discussed thereafter have been realized with unit cells of size $d = 6$~mm which corresponds to $\lambda_0/5$ at 10~GHz. The dielectric substrates used are Rogers RO3003 ($\epsilon_r = 3,~\tan{\delta}=0.001$) with a thickness of 1.52~mm for each spacer leading to a total metasurface thickness of 3.04~mm ($\approx\lambda_0/10$).

\section{Applications}

\begin{figure}[h!]
\centering
\includegraphics[width=1\columnwidth]{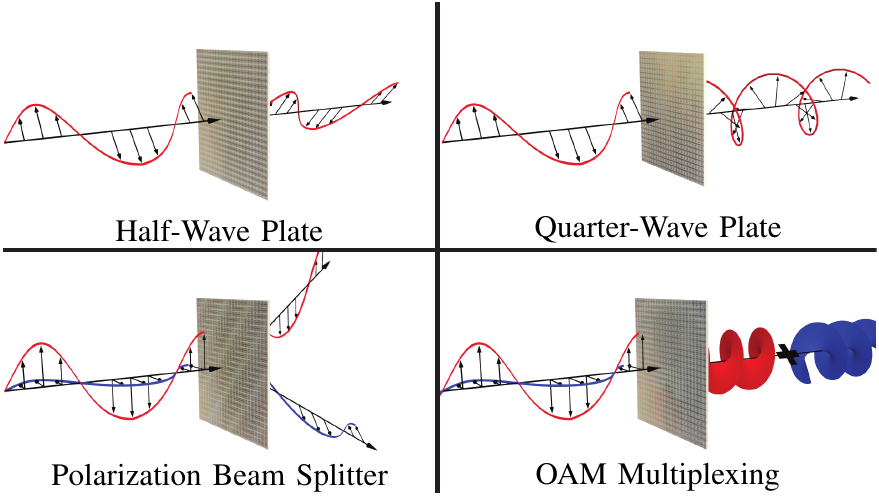}
\caption{Representations of the operation of the four realized metasurfaces.}
\label{Fig:Blend}
\end{figure}

In this section, several types of birefringent metasurfaces are discussed and demonstrated experimentally. The realized metasurface are, in order of appearance, a half-wave plate, a quarter-wave plate, a polarization beam splitter and an orbital angular momentum generator. These four operations are illustrated in Fig.~\ref{Fig:Blend}.

\subsection{Electromagnetic Wave Plates}

Electromagnetic wave plates are birefringent structures that exhibit specific transmission phase difference between $x$- and $y$-polarizations defined as $\Delta\phi= |\phi_x - \phi_y|$. Here, we present the two most common wave plates: a half-wave plate and a quarter-wave plate which, respectively, correspond to $\Delta\phi = \pi$ and $\Delta\phi = \pi/2$. The half-wave plate performs a $90^\circ$ rotation of polarization for linear polarization or changes the handedness of circular polarization. The quarter-wave plate transforms linear polarization into circular polarization and vice-versa.

For such electromagnetic transformations, the metasurface is uniform since there is no variation in the direction of propagation of the waves and, therefore, no phase gradient is required. This make these metasurfaces very easy to design since only one unit cell has to be implemented and repeated periodically to form the metasurface.\\

\subsubsection{Half-Wave Plate}

\begin{figure}[h!]
\centering
\includegraphics[width=0.6\columnwidth]{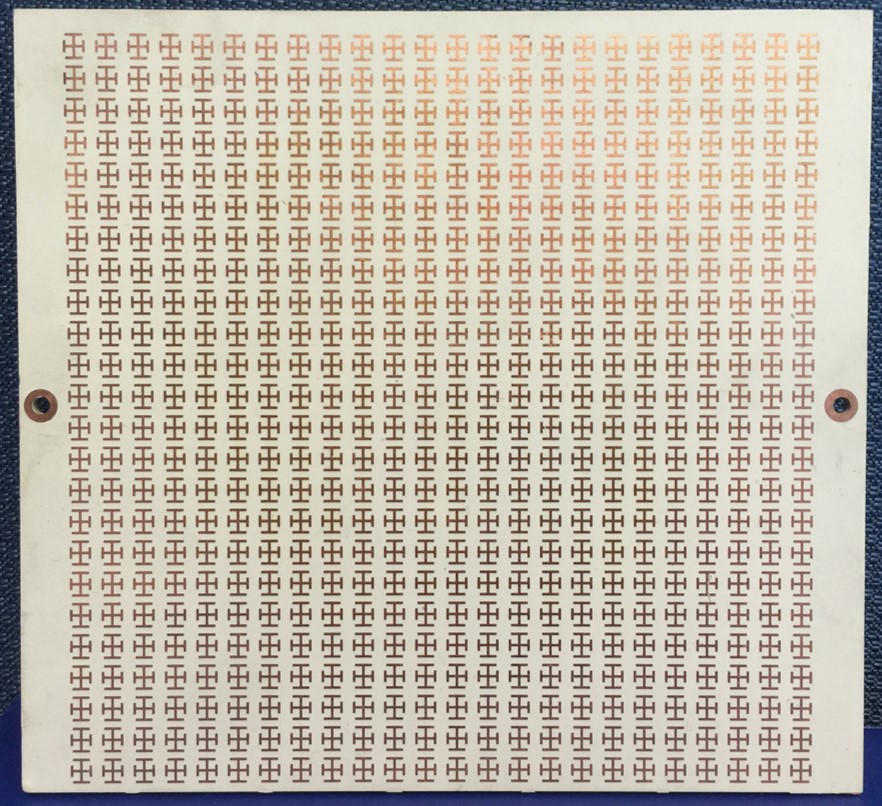}
\caption{Fabricated half-wave plate metasurface.}
\label{Fig:HWP_fab}
\end{figure}

The fabricated metasurface is shown in Fig.~\ref{Fig:HWP_fab} while the dimensions of its unique unit cell are given in Table~\ref{Table1}. The metasurface is made of $24\times24$ unit cells corresponding to an aperture of about 5$\lambda_0\times5\lambda_0$. Note that the two holes on both sides of the metasurface are used to attach it to the measurement setup.
\begin{table}[h!]
\centering
\begin{tabular}{cccccccccc}  \toprule
        & Lx & Ly   & Wx  & Wy   & Ax   & Ay   & Bx   & By    \\\midrule
OL      & 4  & 4.75 & 0.5 & 0.5  & 0.5  & 0.5  & 3.25 & 2.25  \\
ML      & 5  & 5    & 0.5 & 0.25 & 0.25 & 0.75 & 2.25 & 3.5   \\\bottomrule
\end{tabular}
\caption{Geometrical dimensions (in mm) for the unit cell of the metasurface in Fig.~\ref{Fig:HWP_fab}. OL denotes the outer layers and ML the middle layer.}\label{Table1}
\end{table}

\begin{figure}[h!]
\centering
\subfloat[]{\label{Fig:HWP_T1}
\includegraphics[width=0.8\columnwidth]{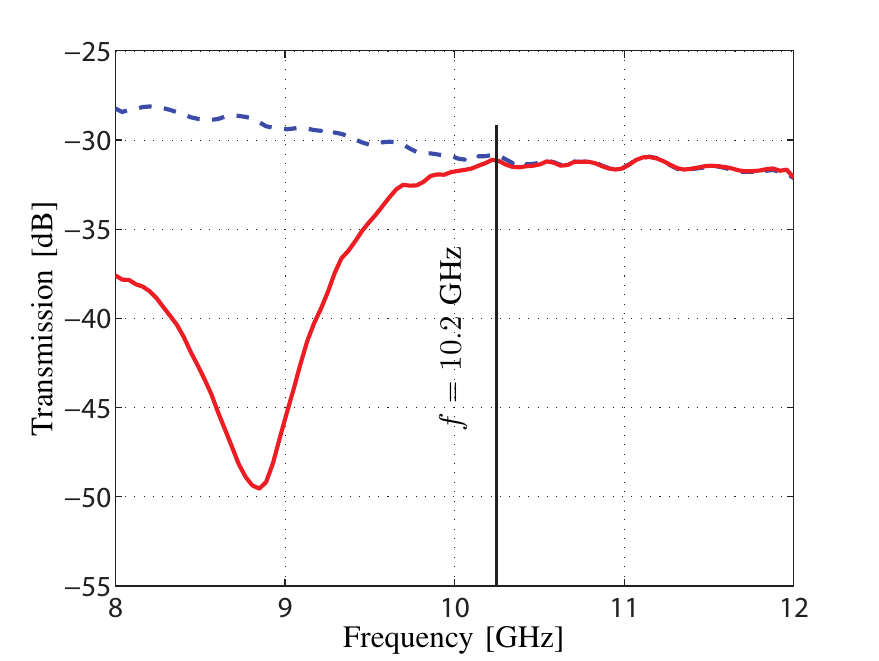}
}\\
\subfloat[]{\label{Fig:HWP_T2}
\includegraphics[width=0.8\columnwidth]{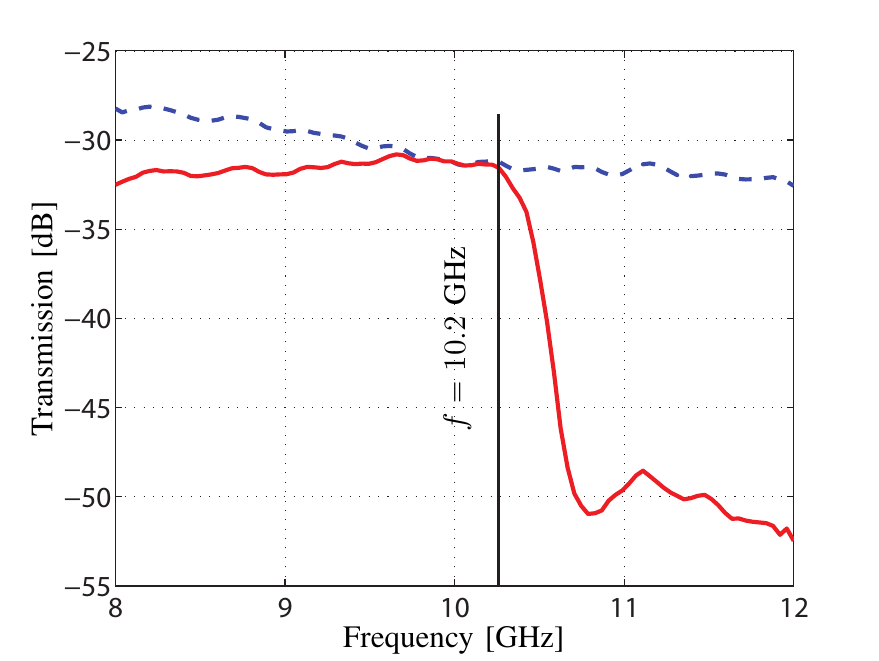}
}\\
\subfloat[]{\label{Fig:HWP_T3}
\includegraphics[width=0.8\columnwidth]{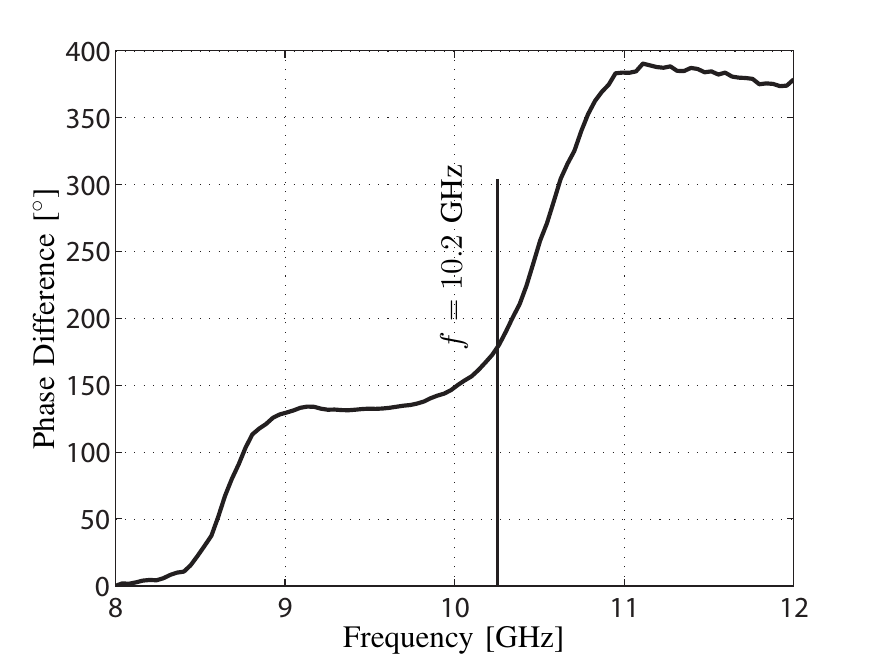}
}
\caption{Measured half-wave plate transmitted power for (a)~$x$-polarization and (b)~$y$-polarization, with phase difference in~(c). In (a) and (b), the blue dashed lines correspond to the reference horn antenna without metasurface while the red lines correspond to the transmission of the metasurface.}
\label{Fig:HWP_T}
\end{figure}

The metasurface has then been measured. Two horn antennas, placed on both sides of the metasurface, have been used to measure the normal transmission from a normally incident wave. The measured transmissions for $x$ and $y$ polarizations are, respectively, plotted in Figs.~\ref{Fig:HWP_T1} and~\ref{Fig:HWP_T2}, where the red solid lines correspond to measurements with the metasurface and the blue dashed lines are the reference line of sight measurements of the horn antennas. The phase difference, $\Delta \phi$, is plotted in Fig~\ref{Fig:HWP_T3}.

\begin{figure}[h!]
\centering
\label{Fig:susc1}
\includegraphics[width=0.8\linewidth]{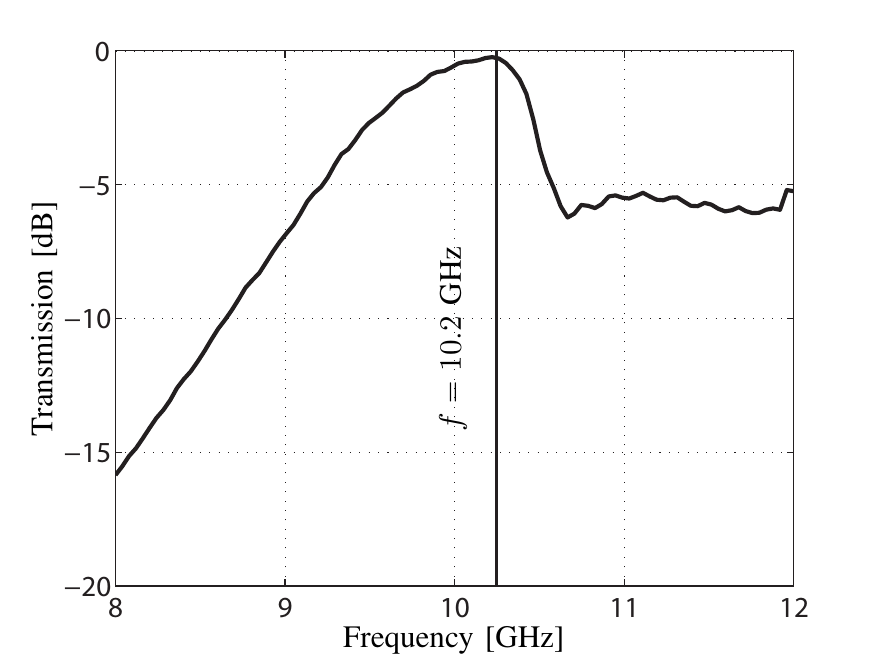}
\caption{Measured and normalized cross-polarized transmission and bandwidth of the half-wave plate.}
\label{Fig:HWP_P}
\end{figure}

As can be seen, the metasurface transmission is almost unity around the specified frequency of operation of 10~GHz. The ideal phase shift difference of $\Delta\phi = \pi$ is obtained at $f=10.2$~GHz. In order to verify the $90^\circ$ rotation of polarization capability of this structure, cross-polarization measurements with and without the metasurface have been performed and the result is shown in Fig.~\ref{Fig:HWP_P} only for the metasurface transmission case for convenience.

The result in Fig.~\ref{Fig:HWP_P} confirms that the metasurface behaves almost as a perfect half-wave plate with a power transmission efficiency of $95\%$ at 10.2~GHz and a -3-dB bandwidth of about $10\%$.\\

\subsubsection{Quarter-Wave Plate}

\begin{figure}[h!]
\centering
\includegraphics[width=0.6\columnwidth]{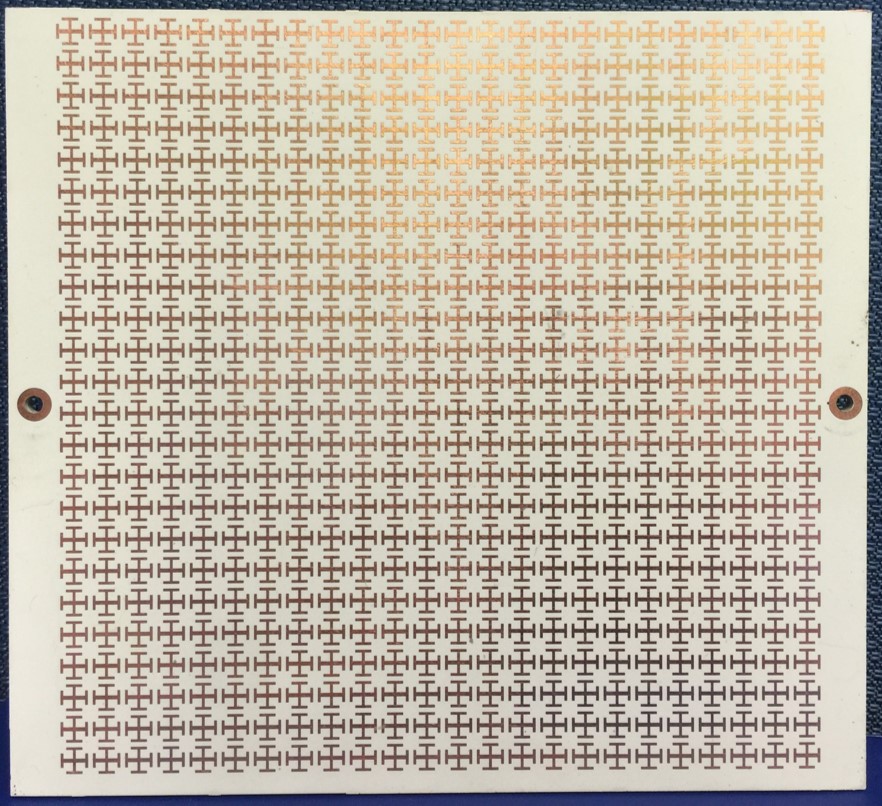}
\caption{Fabricated quarter-wave plate metasurface.}
\label{Fig:QWP_fab}
\end{figure}

The quarter-wave plate metasurface was designed and realized following exactly the same procedure as that of the half-wave plate metasurface. The fabricated metasurface is shown in Fig.~\ref{Fig:QWP_fab} and the dimensions of its unit cell are given in Table~\ref{Table2}.
\begin{table}[h!]
\centering
\begin{tabular}{cccccccccc}  \toprule
        & Lx   & Ly & Wx  & Wy   & Ax   & Ay   & Bx   & By   \\\midrule
OL      & 5    & 5  & 0.5 & 0.75 & 0.5  & 0.5  & 3    & 2.75 \\
ML      & 2.75 & 5  & 0.5 & 0.75 & 0.25 & 0.5  & 2.75 & 1.75 \\\bottomrule
\end{tabular}
\caption{Geometrical dimensions (in mm) for the unit cell of the metasurface in Fig.~\ref{Fig:QWP_fab}. OL denotes the outer layers and ML the middle layer.}\label{Table2}
\end{table}

\begin{figure}[h!]
\centering
\subfloat[]{\label{Fig:QWP_T1}
\includegraphics[width=0.8\columnwidth]{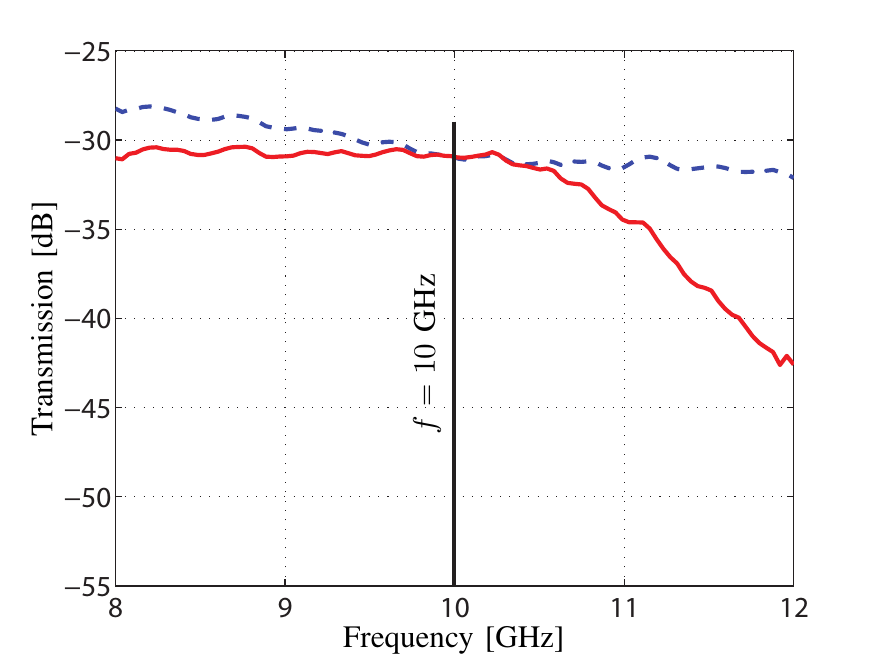}
}\\
\subfloat[]{\label{Fig:QWP_T2}
\includegraphics[width=0.8\columnwidth]{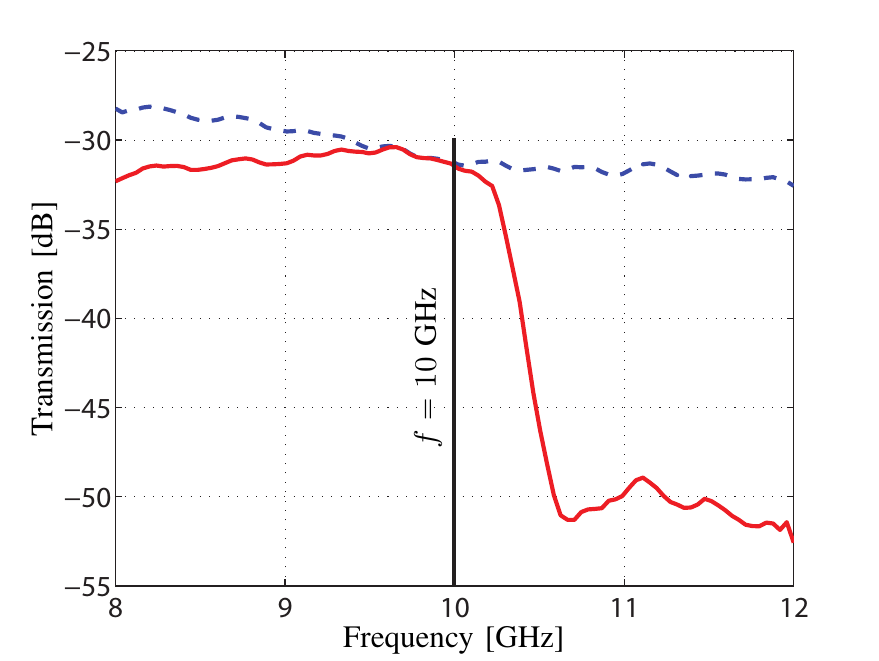}
}\\
\subfloat[]{\label{Fig:QWP_T3}
\includegraphics[width=0.8\columnwidth]{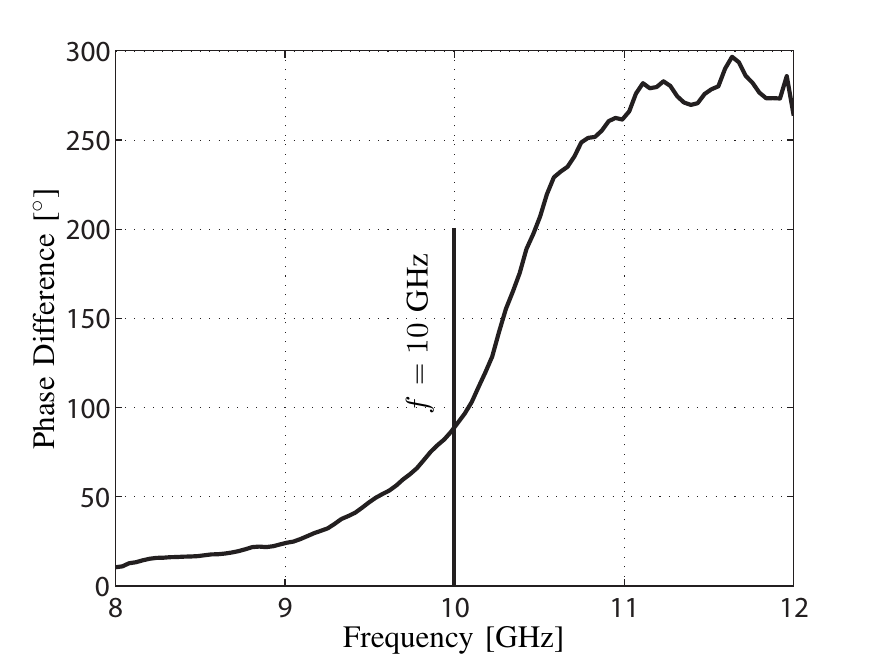}
}
\caption{Measured quarter-wave plate transmitted power for (a) $x$-polarization and (b) $y$-polarization, with phase difference in~(c). In (a) and (b), the blue dashed lines correspond to the reference horn antenna without metasurface while the red lines correspond to the transmission of the metasurface.}
\label{Fig:QWP_T}
\end{figure}

As was done for the half-wave plate metasurface, the measurements of the quarter-wave plate metasurface, corresponding to $x$ polarization transmission, $y$ polarization transmission and phase difference, are plotted in Figs.~\ref{Fig:QWP_T1},~\ref{Fig:QWP_T2} and~\ref{Fig:QWP_T3}, respectively.

\begin{figure}[h!]
\centering
\label{Fig:susc1}
\includegraphics[width=0.8\linewidth]{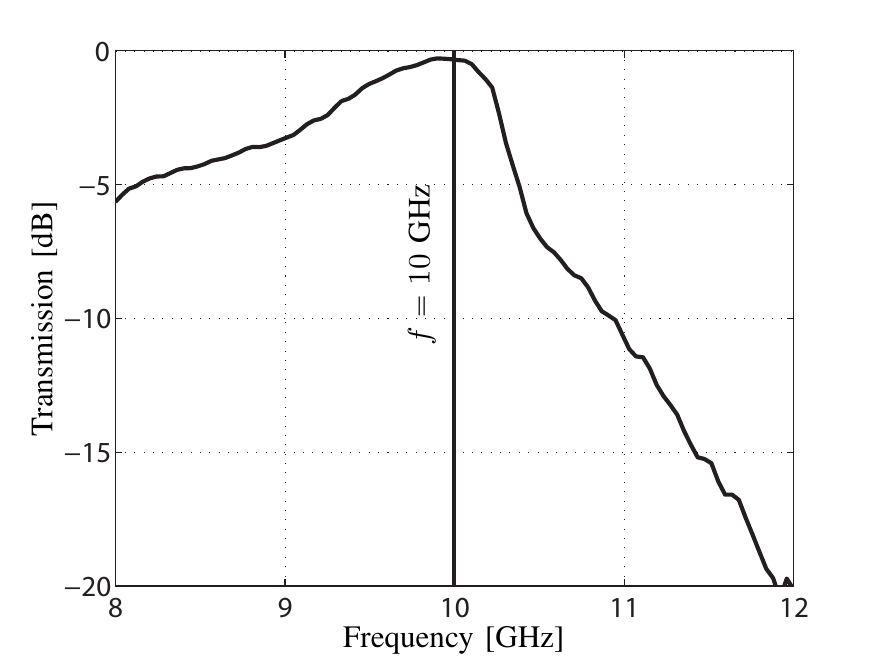}
\caption{Measured linear-to-circular power conversion and bandwidth of the quarter-wave plate.}
\label{Fig:QWP_P}
\end{figure}

The metasurface exhibits very good transmission (near unity) for both $x$ and $y$ polarization around the frequency of operation. The phase difference reaches the required value of $\Delta\phi=\pi/2$ at the specified frequency of 10~GHz. Finally, the linear-to-circular conversion efficiency has been estimated from the $x$ and $y$ polarization amplitude and phase measurements and has been plotted in Fig.~\ref{Fig:QWP_P}.

As can be seen, the linear-to-circular conversion efficiency is very good reaching about $94\%$ at 10~GHz with a -3-dB bandwidth of about $12\%$.\\

\subsection{Generalized Birefringent Reflection and Refraction}

The concept of generalized birefringent reflection and refraction consists in controlling independently the reflection and transmission angles and amplitudes of orthogonally polarized waves. To illustrate this concept, we have realized a polarization beam splitting (PBS) reflection-less metasurface that refracts in opposite directions normally incident $x$ and $y$ polarized waves. The synthesis of this metasurface essentially follows the procedure elaborated in the introduction of this paper and which corresponds to the second synthesis technique that was discussed. Accordingly, the transmission coefficients for $x$ and $y$ polarization are, respectively, given by $T_x(x,y)=e^{-jk_0\sin{\theta_\text{t}}x}$ and $T_y(x,y)=e^{jk_0\sin{\theta_\text{t}}x}$, where $\theta_\text{t}$ is the specified transmission angle. Note that separation of both polarizations was initially specified to be along the $x$ direction. The phase gradient, corresponding to $T_x$ and $T_y$, has a period that is given by $P=\lambda_0/\sin{\theta_\text{t}}$. For the realization of that metasurface, we decided to sample the period $P$ with 8 unit cells of lateral size $\lambda_0/5$. Consequently, the transmission angle is determined by the unit cell size and the number of unit cells and is, thus, given by $\theta_\text{t}=\arcsin{(5/8)}\approx 38.7^\circ$.

Each unit cell has then been implemented to realize a specific phase shift for $x$ and $y$ polarizations, respectively, $\phi_x$ and $\phi_y$. The transmission phases, for each unit cell, are given in Table~\ref{Table3}. Note that the absolute phase shift of a single unit cell is irrelevant, only the phase shift difference between adjacent unit cells (here $45^\circ$) matters.
\begin{table}[h!]
\centering
\begin{tabular}{cccccccccc}  \toprule
        & 1  & 2 & 3  & 4   & 5 & 6 & 7 & 8 \\\midrule
$\phi_x$ & 0 & 45   & 90  & 135   & 180    & 225  & 270  & 315    \\
$\phi_y$ & 315 & 270 & 225  & 180 & 135  & 90  & 45  & 0   \\\bottomrule
\end{tabular}
\caption{Transmission phase shifts (in degrees) for $x$ and $y$ polarization of the 8 unit cells forming the PBS metasurface.}\label{Table3}
\end{table}
As can be seen in Table~\ref{Table3}, the supercell (formed by the 8 unit cells) has an asymmetric phase progression meaning that the unit cells number 5, 6, 7 and 8 have opposite $x$ and $y$ phase shifts as the unit cells number 4, 3, 2 and 1, respectively. This means that the 4 lasts unit cells are simply the $90^\circ$ rotated version of the first 4 unit cells. Consequently, the realization of this metasurface is greatly simplified since only 4 unit cells need to be implemented.

\begin{figure}[h!]
\centering
\includegraphics[width=0.6\columnwidth]{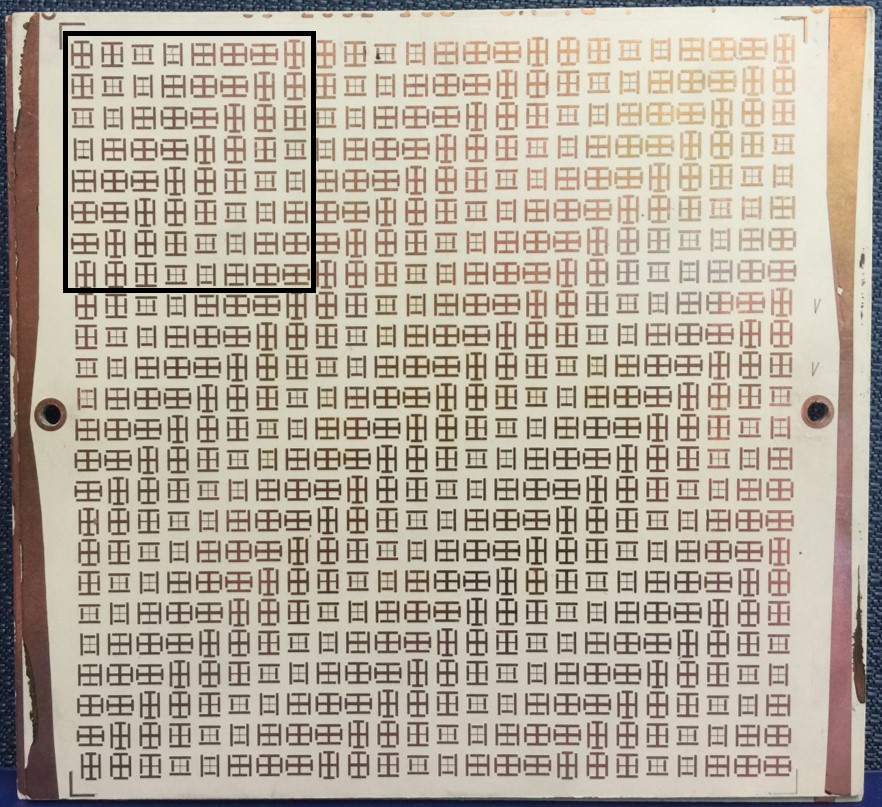}
\caption{Fabricated polarization beam splitting metasurface. The supercell made of $8\times8$ unit cells is highlighted by the black square.}
\label{Fig:PBS_fab}
\end{figure}

After designing the supercell, we performed full-wave simulations to verify the beam splitting behavior of the metasurface. The $x$ polarization refraction yielded good result but, unfortunately, the $y$ polarization transformation was not good, which can be explained by the presence of spurious coupling between adjacent unit cells. While the coupling affected both polarizations, it turns out that it was more damaging to the $y$ polarization than the $x$ polarization. It has to be noted that the metasurface is non-uniform only in the $x$ direction while being perfectly uniform in the $y$ direction, this asymmetry in the structure was hypothesized to be the cause of the different behavior of the two polarizations. To overcome this difficulty, we modified the metasurface such that the same non-uniformity was present in both $x$ and $y$ directions. Consequently, the supercell is now made of $8\times8$ unit cells instead of $8\times1$. The realized metasurface is shown in Fig.~\ref{Fig:PBS_fab}, note the sinusoidally varying pattern in the diagonal direction indicating the direction of the phase gradients. The metasurface is made was a repetition of 9 supercells.
The dimensions of each of the first 4 unit cells are given in Table~\ref{Table4}. As said above, the 4 remaining unit cell are just the rotated version of the first 4 ones.
\begin{table}[]
\centering
\begin{tabular}{ccccccccccc}\toprule
                        &              & Lx   & Ly   & Wx    & Wy     & Ax   & Ay   & Bx    & By    \\\midrule
\multirow{2}{*}{Cell 1} & OL & 5.5  & 4    & 0.5   & 0.625  & 0.5  & 0.5  & 4.875 & 2.25  \\
                        & ML & 5.5  & 5.5  & 0.375 & 1.125  & 0.5  & 0.5  & 1.375 & 2.875 \\\midrule
\multirow{2}{*}{Cell 2} & OL & 5.25 & 4.25 & 0.625 & 0.5    & 0.5  & 0.5  & 4.125 & 3     \\
                        & ML & 5.75 & 3.5  & 0.5   & 0.375  & 0.5  & 0.5  & 0.5   & 1.375 \\\midrule
\multirow{2}{*}{Cell 3} & OL & 4.25 & 4.75 & 0.625 & 0.25   & 0.5  & 0.5  & 4.25  & 3     \\
                        & ML & 3.75 & 5.25 & 0.5   & 0.375  & 0.5  & 0.5  & 2.25  & 4.5   \\\midrule
\multirow{2}{*}{Cell 4} & OL & 3.75 & 3.5  & 0.125 & 0.125  & 0.5  & 0.5  & 2.25  & 4.375 \\
                        & ML & 5.5  & 5.25 & 0.375 & 1.125  & 0.5  & 0.5  & 4.75  & 0.125 \\\bottomrule
\end{tabular}
\caption{Geometrical dimensions (in mm) for the first four unit cells of the metasurface in Fig.~\ref{Fig:PBS_fab}. OL denotes the outer layers and ML the middle layers.}\label{Table4}
\end{table}

Because the metasurface has now a period in the diagonal direction, the dimension of the phase gradient period is reduced to $P=8\lambda_0/(5\sqrt{2})$. This changes the transmission angle to $\theta_\text{t}\approx 62.1^\circ$ at 10~GHz.

\begin{figure}[h!]
\centering
\includegraphics[width=0.8\linewidth]{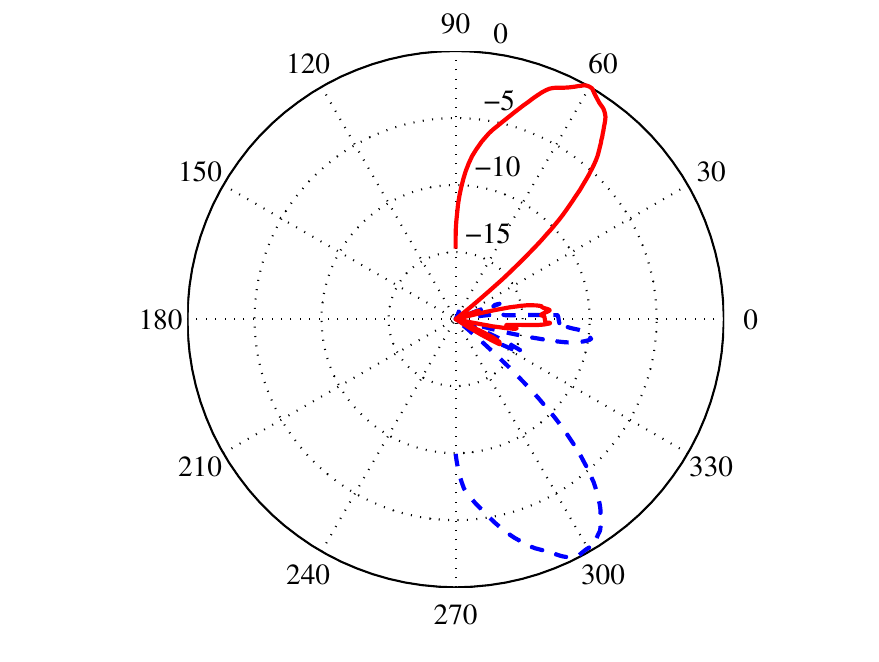}
\caption{Measured normalized transmitted power for $x$-polarization (dashed blue curve) and $y$-polarization.}
\label{Fig:PBS_FF}
\end{figure}

The metasurface has then been measured. A horn antenna was used to generate the normally incident waves while a probe was scanning the near-field over a plane parallel to the metasurface in the transmission region. Near-field to far-field transformation~\cite{balanis2016antenna} was then used to evaluate the transmission response of the metasurface. The measured $x$ and $y$ polarization transmissions, in the diagonal plane of the metasurface, are plotted in Fig.~\ref{Fig:PBS_FF} as a dashed blue line and a solid red line, respectively. Note that the curves have been normalized with respect to the $y$-polarized transmission.

\begin{figure}[h!]
\centering
\includegraphics[width=0.8\linewidth]{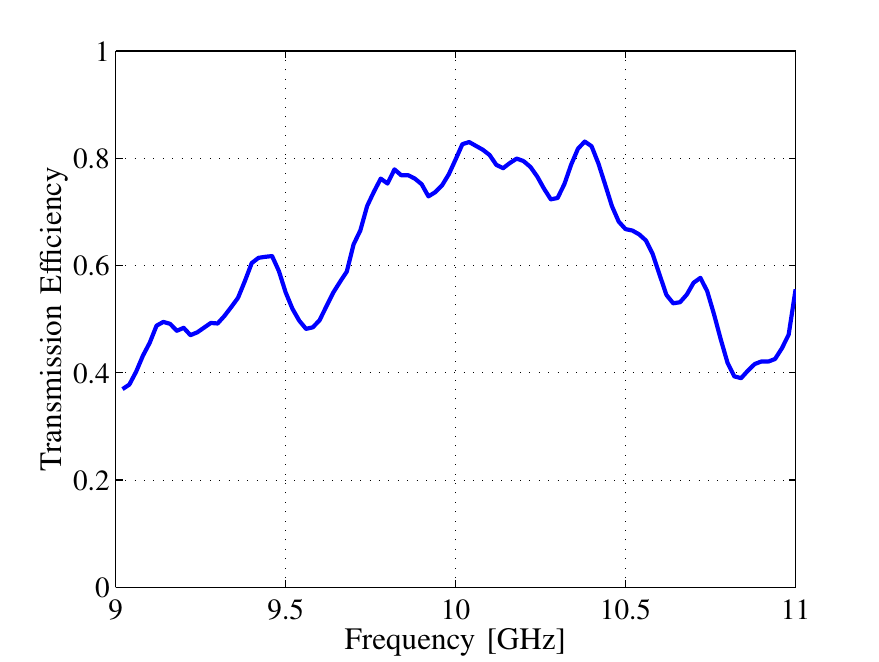}
\caption{PBS metasurface transmission efficiency.}
\label{Fig:PBS_Eff}
\end{figure}

As can be seen, the metasurface effectively separates the two polarizations which are refracted, with almost identical amplitude, at about $+60^\circ$ and $-60^\circ$ from broadside, respectively. The frequency corresponding to the results in Fig.~\ref{Fig:PBS_FF} is about 10.4~GHz and the transmission efficiency, defined has the ratio between the transmitted electric field and the incident electric field, is about $80\%$. The efficiency of the metasurface versus frequency is plotted in Fig.~\ref{Fig:PBS_Eff}. The reasons for which the metasurface efficiency does not exceed $80\%$ can be explained partly by the presence of loss in the dielectric layers but mostly from undesired refraction orders (either in reflection or in transmission) that are due to the spurious coupling of the unit cells. For instance, zeroth diffraction orders are clearly visible in the measurements shown in Fig.~\ref{Fig:PBS_FF}.

\subsection{Orbital Angular Momentum Multiplexing}

The last metasurface that was realized is a structure that generates waves possessing orbital angular momentum (OAM) of different topological charges depending on the polarization of the incident wave. The OAM wave that we have chosen as the transmitted wave is a Hypergeometric Gaussian (HyG) wave that corresponds to the solution of the paraxial Maxwell equations in cylindrical coordinates. The reason why the HyG wave has been chosen over the more common Bessel wave~\cite{recami2009localized} is that the HyG has the advantage of being linearly polarized whereas the Bessel wave is either radially of longitudinally polarized making the multiplexing of two OAM waves with a single metasurface more difficult for Bessel waves.

The metasurface is thus required to transform an $x$-polarized normally incident wave into an HyG wave of topological charge $m=+1$ and $y$-polarized normally incident wave into an HyG wave of topological charge $m=-1$. The electric field of an HyG wave reads~\cite{karimiHYG}
\begin{equation}
\begin{split}
\label{eq:HYG}
E(\rho,\phi,z) =& \frac{\Gamma \left ( 1+|m|+\frac{p}{2}  \right )}{\Gamma (|m|+1)}\frac{i^{|m|+1}\zeta^{|m|/2}\xi^{p/2}}{[\xi+i]^{1+|m|/2+p/2}}e^{im\phi-i\zeta}\\
&\times~_{1}F_1 \left ( -\frac{p}{2},|m|+1;\frac{\zeta[\xi + i]}{\xi[\xi - i]} \right ),
\end{split}
\end{equation}
where $_{1}F_1(a,b;x)$ is the confluent hypergeometric function, $\Gamma(x)$ is the gamma function, $m$ is the OAM topological charge, $p\geqslant -|m|$ is a real parameter, and where $\zeta = \rho^2/(w_0^2[\xi+i])$, $\xi=z/z_R$ $w_0$, with $w_0$ being the beam waist and $z_r$ the Rayleigh range given by $z_r=\pi w_0^2/\lambda$.

\begin{figure}[h!]
\centering
\subfloat[]{\label{Fig:Hyg1}
\includegraphics[width=0.5\linewidth]{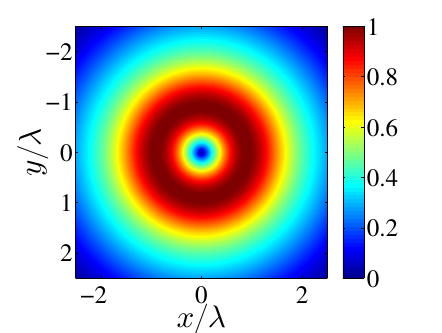}
}
\subfloat[]{\label{Fig:Hyg2}
\includegraphics[width=0.5\linewidth]{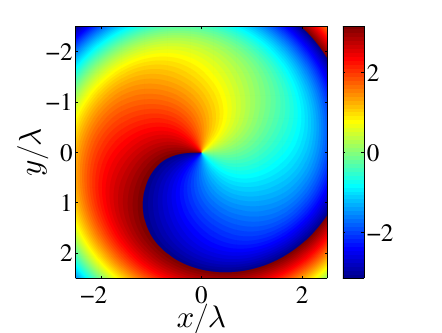}
}\\
\subfloat[]{\label{Fig:Hyg3}
\includegraphics[width=0.5\linewidth]{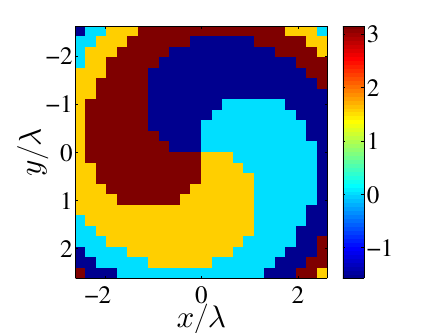}
}
\subfloat[]{\label{Fig:Hyg4}
\includegraphics[width=0.5\linewidth]{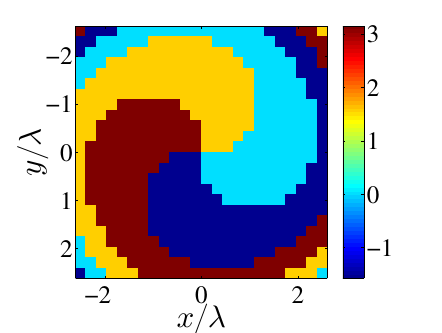}
}
\caption{Hypergeometric Gaussian wave (a)~amplitude and (b)~phase for parameters $p=1$, $m=-1$, $w_0=\lambda$ and $\xi=1$. The metasurface simplified phase patterns for $x$ and $y$ polarizations is given in (c) and (d) for topological charges $m=+1$ and $m=-1$, respectively.}
\label{Fig:Hyg}
\end{figure}

The amplitude and phase of the HyG wave, for $m=-1$, are plotted in Figs.~\ref{Fig:Hyg1} and~\ref{Fig:Hyg2}, respectively. It is clear that this kind of wave has a non-periodic phase pattern, as evidenced in Fig.~\ref{Fig:Hyg2}, contrary to the oblique transmitted plane waves that were specified for the polarization beam splitting metasurface in the previous section. This means that a larger number of unit cells has to be implemented because of the aperiodicity of the transformation. Moreover, the amplitude of the HyG wave is non-uniform which further complicates the realization of the unit cells. However, these difficulties may by overcome by assuming that the amplitude of the $x$ and $y$ transmission coefficients are $|T_x(x,y)| = |T_y(x,y)|=1$ instead of following the profile in Fig.~\ref{Fig:Hyg1}. Despite the fact that this approximation might a priori seem extreme, it turns out that the main properties of the HyG wave may be obtained by only implementing its phase evolution. For instance, the null amplitude at the center of the wave is achieved by destructive interferences due to the phase rotation around the center. Moreover, the orbital angular momentum information is contained not in the amplitude but rather in the phase of the wave. These considerations justify the assumption that only the phase profile of the transmitted waves should be implemented while their respective magnitude can be assumed to be uniform and equal to 1. Additionally, the phase profiles of the two OAM waves were discretized by four phase samples each. Consequently, the metasurface is made of $24\times 24$ unit cells with phase shifts for $x$ and $y$ polarizations as given in Figs.~\ref{Fig:Hyg3} and~\ref{Fig:Hyg4}, respectively.

\begin{figure}[h!]
\centering
\includegraphics[width=0.7\linewidth]{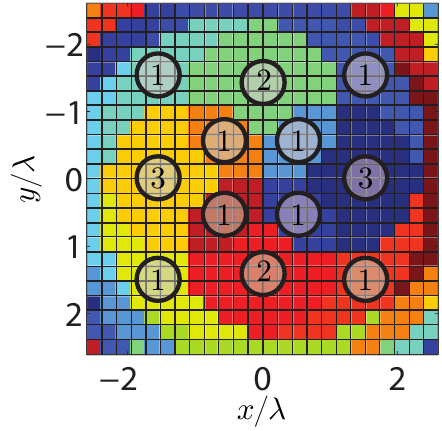}
\caption{Representation of the $24\times 24$ unit cells of the OAM multiplexing metasurface. Each color represents a unit cell with specific phase shifts for $x$ and $y$ polarizations. There is a total number of 16 different unit cells where, notably, the ones in regions 1 are quarter-wave plates, the ones in regions 2 are half-wave plates and the ones in regions 3 are isotropic.}
\label{Fig:Hyg_comb}
\end{figure}

\begin{figure}[h!]
\centering
\includegraphics[width=0.6\columnwidth]{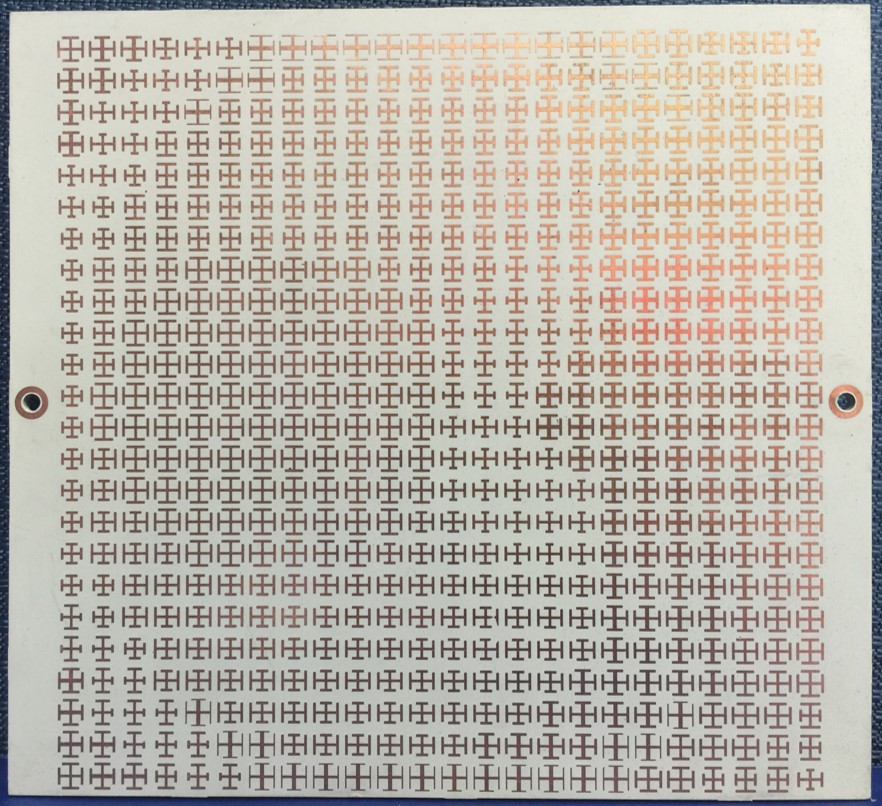}
\caption{Fabricated orbital angular momentum multiplexing metasurface. It is made of 16 uniquely different unit cells arranged according to the pattern in Fig.~\ref{Fig:Hyg_comb}, the metasurface contains a total of $24\times24$ unit cells.}
\label{Fig:HYG_fab}
\end{figure}

When combining together the phase shifts in Figs.~\ref{Fig:Hyg3} and~\ref{Fig:Hyg4}, it turns out that the total number of different unit cells composing the metasurface is 16. The Fig.~\ref{Fig:Hyg_comb} represents the $24\times 24$ unit cells of the metasurface. In that figure, each color corresponds to a specific unit cell having unique phase shift for $x$ and $y$ polarizations. Interestingly, the unit cells in the highlighted regions 1, 2 and 3 are quarter-wave plates, half-wave plates and isotropic wave plates (where $\phi_x=\phi_y=\phi$), respectively. Consequently, 6 out of the 10 correspond to the same unit cells but rotated in the plane of the metasurface with respect to each other.
The dimensions for the 10 remaining unit cells are given in Table~\ref{Table5} and the fabricated metasurface is shown in Fig.~\ref{Fig:HYG_fab}.
\begin{table}[]
\centering
\begin{tabular}{ccccccccccc}\toprule
                         &    & Lx   & Ly   & Wx   & Wy   & Ax   & Ay   & Bx   & By   \\\midrule
\multirow{2}{*}{(45,45)}  & OL & 5    & 5    & 0.5  & 0.5  & 0.5  & 0.5  & 3.5  & 3.5  \\
                         & ML & 4.75 & 4.75 & 0.75 & 0.75 & 0.5  & 0.5  & 3.5  & 3.5  \\\midrule
\multirow{2}{*}{(45,135)}  & OL & 5    & 4.75 & 0.5  & 0.5  & 0.5  & 0.5  & 3.25 & 3.25 \\
                         & ML & 4.75 & 4.75 & 0.25 & 1    & 1    & 0.5  & 2.5  & 3.75 \\\midrule
\multirow{2}{*}{(45,225)}  & OL & 4.75 & 5    & 0.75 & 0.5  & 0.25 & 0.25 & 2.75 & 4.25 \\
                         & ML & 4.5  & 2.75 & 0.25 & 0.5  & 0.5  & 0.75 & 1.75 & 3.25 \\\midrule
\multirow{2}{*}{(45,315)}  & OL & 4.75 & 4    & 0.5  & 0.5  & 0.5  & 0.5  & 2.25 & 3.75 \\
                         & ML & 4.75 & 5    & 0.25 & 0.5  & 0.5  & 0.5  & 3    & 2.75 \\\midrule
\multirow{2}{*}{(135,135)}  & OL & 4.75 & 4.75 & 0.5  & 0.5  & 0.5  & 0.5  & 3.25 & 3.25 \\
                         & ML & 4.75 & 4.75 & 0.5  & 0.5  & 0.75 & 0.75 & 2.75 & 2.75 \\\midrule
\multirow{2}{*}{(135,225)}  & OL & 4.75 & 5    & 0.5  & 0.75 & 0.5  & 0.5  & 3    & 3    \\
                         & ML & 2.75 & 4.5  & 0.5  & 0.75 & 0.25 & 0.75 & 3.25 & 1.75 \\\midrule
\multirow{2}{*}{(135,315)}  & OL & 4    & 4.75 & 0.5  & 0.5  & 0.5  & 0.5  & 3.25 & 2.25 \\
                         & ML & 5    & 4.75 & 0.5  & 0.25 & 0.5  & 0.75 & 2.5  & 3    \\\midrule
\multirow{2}{*}{(225,225)}  & OL & 4.75 & 4.75 & 0.75 & 0.75 & 0.5  & 0.5  & 3.25 & 3.25 \\
                         & ML & 2.75 & 2.75 & 0.75 & 0.75 & 0.25 & 0.25 & 1.75 & 1.75 \\\midrule
\multirow{2}{*}{(225,315)}  & OL & 4    & 4.75 & 0.5  & 0.5  & 0.5  & 0.5  & 2.75 & 1.5  \\
                         & ML & 5    & 2.75 & 0.75 & 0.25 & 0.5  & 0.25 & 1.75 & 4.25 \\\midrule
\multirow{2}{*}{(315,315)} & OL & 4    & 4    & 0.5  & 0.5  & 0.5  & 0.5  & 2.25 & 2.25 \\
                         & ML & 5    & 5    & 0.25 & 0.25 & 0.5  & 0.5  & 3    & 3    \\\bottomrule
\end{tabular}
\caption{Geometrical dimensions (in mm) for 10 of the unit cells of the metasurface in Fig.~\ref{Fig:HYG_fab}. OL denotes the outer layers and ML the middle layers. The numbers in the first column correspond to the phase shifts $(\phi_x,\phi_y)$ for $x$ and $y$ polarizations, respectively.}\label{Table5}
\end{table}

\begin{figure}[h!]
\centering
\subfloat[]{
\includegraphics[width=0.5\linewidth]{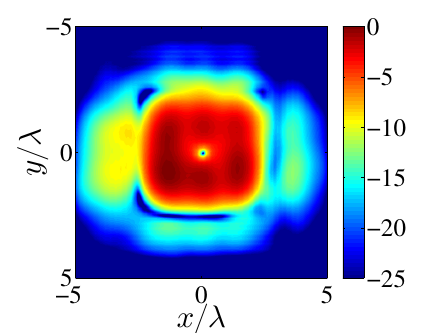}
}
\subfloat[]{
\includegraphics[width=0.5\linewidth]{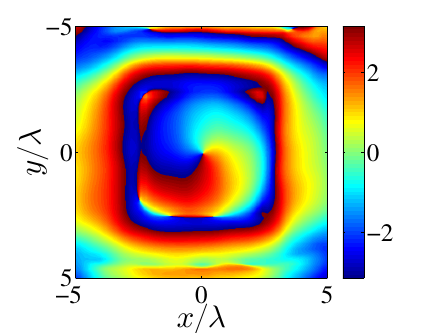}
}\\
\subfloat[]{
\includegraphics[width=0.5\linewidth]{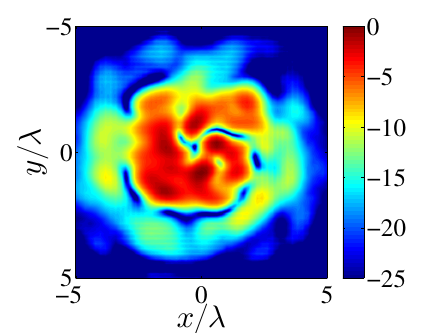}
}
\subfloat[]{
\includegraphics[width=0.5\linewidth]{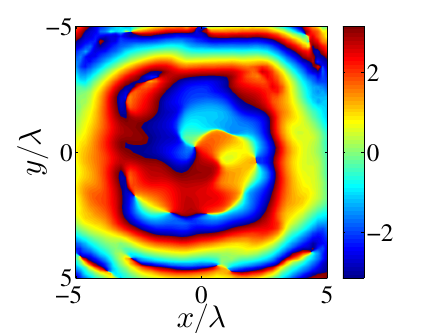}
}
\caption{$x$-polarized simulated (a)~amplitude and (b)~phase of the expected metasurface scattered field by taking into account the radiation of the horn antenna. Corresponding measured (c)~amplitude and (d)~phase of the metasurface scattered field.}
\label{Fig:Hyg_aclk}
\end{figure}

\begin{figure}[h!]
\centering
\subfloat[]{
\includegraphics[width=0.5\linewidth]{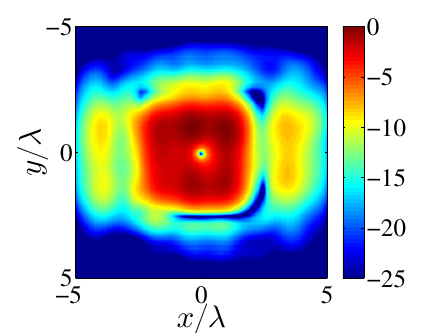}
}
\subfloat[]{
\includegraphics[width=0.5\linewidth]{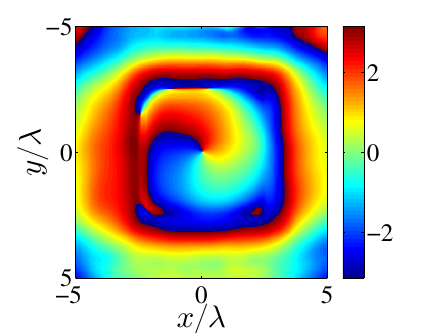}
}\\
\subfloat[]{
\includegraphics[width=0.5\linewidth]{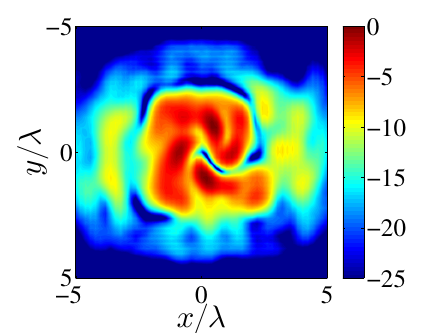}
}
\subfloat[]{
\includegraphics[width=0.5\linewidth]{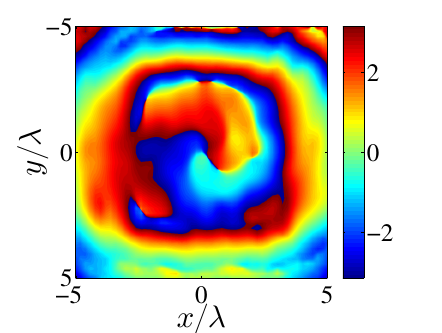}
}
\caption{$y$-polarized simulated (a)~amplitude and (b)~phase of the expected metasurface scattered field by taking into account the radiation of the horn antenna. Corresponding measured (c)~amplitude and (d)~phase of the metasurface scattered field.}
\label{Fig:Hyg_clk}
\end{figure}

The metasurface has then been simulated and measured and the results are reported in Figs.~\ref{Fig:Hyg_aclk} and~\ref{Fig:Hyg_clk} which respectively correspond to $x$ and $y$ polarizations. In these two figures, plots (a) and (b) are the simulated amplitude and phase transmissions. These simulations were obtained by first measuring the radiated reference field of the exciting horn antenna at the position of the metasurface. Then, the expected scattered field of the metasurface was calculated using the antenna reference field and assuming ideal transmission of flat and unity amplitude and phase profiles as in Figs.~\ref{Fig:Hyg3} and~\ref{Fig:Hyg4}. This simulation technique allows fair comparison between the expected scattered fields and the measured scattered fields from the metasurface that are shown in plots (c) and (d).

\begin{figure}[h!]
\centering
\includegraphics[width=0.8\linewidth]{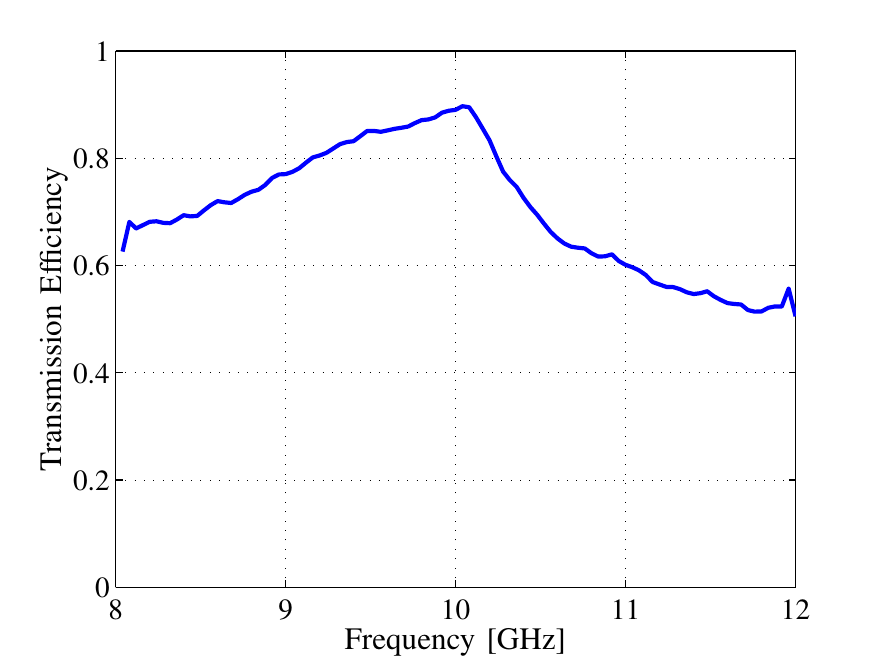}
\caption{Orbital angular momentum multiplexing metasurface transmission efficiency.}
\label{Fig:Hyg_eff}
\end{figure}

The measured results are in good agreements with the expected simulated results. The topological charges of $m=+1$ and $m=-1$ are achieved with a transmission efficiency near $90\%$ at 10~GHz. Finally, the transmission efficiency of the metasurface was evaluated for a frequency band between 8 and 12 GHz. The result is reported in Fig.~\ref{Fig:Hyg_eff}.

\section{Conclusion}

In the first part of this work, we have discussed and elaborated two different approaches for the synthesis of birefringent metasurfaces. The two methods yield the metasurface electric and magnetic susceptibilities either when the exact electromagnetic fields on both sides on the metasurface are specified or when the transmission and reflection coefficients are specified. The first synthesis technique is rigorous while the second one is an approximation. However, we have seen that both methods lead to similar results in terms of the metasurface scattering response. One of the main differences between the two techniques lies in the difficulty of the physical realization of the metasurfaces. While the first technique requires the implementation of complicated non-uniform amplitude and phase transmission and reflection coefficients, the second usually only requires the implementation of non-uniform phase profiles. Consequently, although less rigorous, the second synthesis method is generally preferred over the first one.

In the second part, we have presented the synthesis and realization of four different birefringent metasurfaces performing the operation of half-wave plate, quarter-wave plate, polarization beam splitting and orbital angular momentum multiplexing, respectively. These metasurfaces were synthesized based on the second synthesis technique and their measurements were found in good agreement with the expected scattering responses.

\bibliographystyle{IEEEtran}
\bibliography{LIB}

\begin{thebibliography}{10}
\providecommand{\url}[1]{#1}
\csname url@samestyle\endcsname
\providecommand{\newblock}{\relax}
\providecommand{\bibinfo}[2]{#2}
\providecommand{\BIBentrySTDinterwordspacing}{\spaceskip=0pt\relax}
\providecommand{\BIBentryALTinterwordstretchfactor}{4}
\providecommand{\BIBentryALTinterwordspacing}{\spaceskip=\fontdimen2\font plus
\BIBentryALTinterwordstretchfactor\fontdimen3\font minus
  \fontdimen4\font\relax}
\providecommand{\BIBforeignlanguage}[2]{{%
\expandafter\ifx\csname l@#1\endcsname\relax
\typeout{** WARNING: IEEEtran.bst: No hyphenation pattern has been}%
\typeout{** loaded for the language `#1'. Using the pattern for}%
\typeout{** the default language instead.}%
\else
\language=\csname l@#1\endcsname
\fi
#2}}
\providecommand{\BIBdecl}{\relax}
\BIBdecl

\bibitem{saleh2007fundamentals}
B.~Saleh and M.~Teich, \emph{Fundamentals of Photonics}, ser. Wiley Series in
  Pure and Applied Optics.\hskip 1em plus 0.5em minus 0.4em\relax Wiley, 2007.

\bibitem{Bartho}
E.~Bartholin, \emph{Experimenta crystalli islandici disdiaclastici quibus mira
  \& infolita refractio detegitur}.\hskip 1em plus 0.5em minus 0.4em\relax
  Phil. Trans. of the Royal Society of London, 1669.

\bibitem{yu2014flat}
N.~Yu and F.~Capasso, ``Flat optics with designer metasurfaces,'' \emph{Nature
  Mater.}, vol.~13, no.~2, pp. 139--150, 2014.

\bibitem{lin2014dielectric}
D.~Lin, P.~Fan, E.~Hasman, and M.~L. Brongersma, ``Dielectric gradient
  metasurface optical elements,'' \emph{science}, vol. 345, no. 6194, pp.
  298--302, 2014.

\bibitem{Pfeiffer2013a}
C.~Pfeiffer and A.~Grbic, ``Metamaterial {H}uygens' surfaces: tailoring wave
  fronts with reflectionless sheets,'' \emph{Phys. Rev. Lett.}, vol. 110, p.
  197401, May 2013.

\bibitem{023902}
C.~Pfeiffer, C.~Zhang, V.~Ray, L.~J. Guo, and A.~Grbic, ``High performance
  bianisotropic metasurfaces: asymmetric transmission of light,'' \emph{Phys.
  Rev. Lett.}, vol. 113, p. 023902, Jul 2014.

\bibitem{kim2014optical}
M.~Kim, A.~M. Wong, and G.~V. Eleftheriades, ``Optical huygens’ metasurfaces
  with independent control of the magnitude and phase of the local reflection
  coefficients,'' \emph{Physical Review X}, vol.~4, no.~4, p. 041042, 2014.

\bibitem{achouri2014general}
K.~Achouri, M.~A. Salem, and C.~Caloz, ``General metasurface synthesis based on
  susceptibility tensors,'' \emph{IEEE Trans. Antennas Propag.}, vol.~63,
  no.~7, pp. 2977--2991, July 2015.

\bibitem{art1}
------, ``Birefringent generalized refractive metasurface,'' in \emph{2015 IEEE
  International Symposium on Antennas and Propagation USNC/URSI National Radio
  Science Meeting}, July 2015, pp. 878--879.

\bibitem{AchouriEPJAM}
K.~Achouri, B.~A. Khan, S.~Gupta, G.~Lavigne, M.~A. Salem, and C.~Caloz,
  ``Synthesis of electromagnetic metasurfaces: principles and illustrations,''
  \emph{EPJ Applied Metamaterials}, vol.~2, p.~12, 2015.

\bibitem{ding2015broadband}
F.~Ding, Z.~Wang, S.~He, V.~M. Shalaev, and A.~V. Kildishev, ``Broadband
  high-efficiency half-wave plate: a supercell-based plasmonic metasurface
  approach,'' \emph{ACS nano}, vol.~9, no.~4, pp. 4111--4119, 2015.

\bibitem{pors2013broadband}
A.~Pors, M.~G. Nielsen, and S.~I. Bozhevolnyi, ``Broadband plasmonic half-wave
  plates in reflection,'' \emph{Optics letters}, vol.~38, no.~4, pp. 513--515,
  2013.

\bibitem{yu2012broadband}
N.~Yu, F.~Aieta, P.~Genevet, M.~A. Kats, Z.~Gaburro, and F.~Capasso, ``A
  broadband, background-free quarter-wave plate based on plasmonic
  metasurfaces,'' \emph{Nano letters}, vol.~12, no.~12, pp. 6328--6333, 2012.

\bibitem{zhao2011manipulating}
Y.~Zhao and A.~Al{\`u}, ``Manipulating light polarization with ultrathin
  plasmonic metasurfaces,'' \emph{Physical Review B}, vol.~84, no.~20, p.
  205428, 2011.

\bibitem{pors2013gap}
A.~Pors, O.~Albrektsen, I.~P. Radko, and S.~I. Bozhevolnyi, ``Gap plasmon-based
  metasurfaces for total control of reflected light,'' \emph{Scientific
  reports}, vol.~3, 2013.

\bibitem{farmahini2013birefringent}
M.~Farmahini-Farahani and H.~Mosallaei, ``Birefringent reflectarray metasurface
  for beam engineering in infrared,'' \emph{Optics letters}, vol.~38, no.~4,
  pp. 462--464, 2013.

\bibitem{lee2014semiconductor}
J.~H. Lee, J.~W. Yoon, M.~J. Jung, J.~K. Hong, S.~H. Song, and R.~Magnusson,
  ``A semiconductor metasurface with multiple functionalities: A polarizing
  beam splitter with simultaneous focusing ability,'' \emph{Applied Physics
  Letters}, vol. 104, no.~23, p. 233505, 2014.

\bibitem{PhysRevApplied.2.044012}
C.~Pfeiffer and A.~Grbic, ``Controlling vector bessel beams with
  metasurfaces,'' \emph{Phys. Rev. Applied}, vol.~2, p. 044012, Oct 2014.

\bibitem{karimi2014generating}
E.~Karimi, S.~A. Schulz, I.~De~Leon, H.~Qassim, J.~Upham, and R.~W. Boyd,
  ``Generating optical orbital angular momentum at visible wavelengths using a
  plasmonic metasurface,'' \emph{Light: Science \& Applications}, vol.~3,
  no.~5, p. e167, 2014.

\bibitem{capasso1}
N.~Yu, P.~Genevet, M.~A. Kats, F.~Aieta, J.-P. Tetienne, F.~Capasso, and
  Z.~Gaburro, ``Light propagation with phase discontinuities: generalized laws
  of reflection and refraction,'' \emph{Science}, vol. 334, no. 6054, pp.
  333--337, 2011.

\bibitem{chen2015creating}
C.-F. Chen, C.-T. Ku, Y.-H. Tai, P.-K. Wei, H.-N. Lin, and C.-B. Huang,
  ``Creating optical near-field orbital angular momentum in a gold
  metasurface,'' \emph{Nano letters}, vol.~15, no.~4, pp. 2746--2750, 2015.

\bibitem{wang2015ultra}
W.~Wang, Y.~Li, Z.~Guo, R.~Li, J.~Zhang, A.~Zhang, and S.~Qu, ``Ultra-thin
  optical vortex phase plate based on the metasurface and the angular momentum
  transformation,'' \emph{Journal of Optics}, vol.~17, no.~4, p. 045102, 2015.

\bibitem{Idemen1973}
M.~M. Idemen, \emph{Discontinuities in the Electromagnetic Field}.\hskip 1em
  plus 0.5em minus 0.4em\relax John Wiley \& Sons, 2011.

\bibitem{asadchy2016metasurfaces}
V.~Asadchy, M.~Albooyeh, S.~Tcvetkova, Y.~Ra'di, and S.~Tretyakov,
  ``Metasurfaces for perfect and full control of refraction and reflection,''
  \emph{arXiv preprint arXiv:1603.07186}, 2016.

\bibitem{Salem2013c}
M.~A. Salem and C.~Caloz, ``Manipulating light at distance by a metasurface
  using momentum transformation,'' \emph{Opt. Express}, vol.~22, no.~12, pp.
  14\,530--14\,543, Jun 2014.

\bibitem{7274678}
B.~O. Zhu and Y.~Feng, ``Passive metasurface for reflectionless and arbitary
  control of electromagnetic wave transmission,'' \emph{IEEE Transactions on
  Antennas and Propagation}, vol.~63, no.~12, pp. 5500--5511, Dec 2015.

\bibitem{vahabzadeh2016simulation}
Y.~Vahabzadeh, K.~Achouri, and C.~Caloz, ``Simulation of metasurfaces in finite
  difference techniques,'' \emph{arXiv preprint arXiv:1602.04086}, 2016.

\bibitem{PhysRevApplied.2.044011}
C.~Pfeiffer and A.~Grbic, ``Bianisotropic metasurfaces for optimal polarization
  control: Analysis and synthesis,'' \emph{Phys. Rev. Applied}, vol.~2, p.
  044011, Oct 2014.

\bibitem{balanis2016antenna}
C.~A. Balanis, \emph{Antenna theory: analysis and design}.\hskip 1em plus 0.5em
  minus 0.4em\relax John Wiley \& Sons, 2016.

\bibitem{recami2009localized}
E.~Recami and M.~Zamboni-Rached, ``Localized waves: a review,'' \emph{Advances
  in Imaging and Electron Physics}, vol. 156, pp. 235--353, 2009.

\bibitem{karimiHYG}
E.~Karimi, G.~Zito, B.~Piccirillo, L.~Marrucci, and E.~Santamato,
  ``Hypergeometric-gaussian beams,'' \emph{Opt. Lett.}, vol.~32, no.~21, pp.
  3053--3055, 2007.

\end{thebibliography}

\end{document}